# Ultraviolet Shadowing of RNA Causes Substantial Non-Poissonian Chemical Damage in Seconds

Wipapat Kladwang[1], Justine Hum[1,‡], and Rhiju Das[1,2*]

Department of Biochemistry[1] and Department of Physics[2], Stanford University, Stanford, California 94305

[*] To whom correspondence should be addressed. Phone: (650) 723-5976. Fax: (650) 723-6783. E-mail: rhiju@stanford.edu.

[‡] Current address: University of Vermont College of Medicine

**ABSTRACT**

Chemical purity of RNA samples is critical for high-precision studies of RNA folding and catalytic behavior, but such purity may be compromised by photodamage accrued during ultraviolet (UV) visualization of gel-purified samples. Here, we quantitatively assess the breadth and extent of such damage by using reverse transcription followed by single-nucleotide-resolution capillary electrophoresis. We detected UV-induced lesions across a dozen natural and artificial RNAs including riboswitch domains, other non-coding RNAs, and artificial sequences; across multiple sequence contexts, dominantly at but not limited to pyrimidine doublets; and from multiple lamps that are recommended for UV shadowing in the literature. Most strikingly, irradiation time-courses reveal detectable damage within a few seconds of exposure, and these data can be quantitatively fit to a 'skin effect' model that accounts for the increased exposure of molecules near the top of irradiated gel slices. The results indicate that 200-nucleotide RNAs subjected to 20 seconds or less of UV shadowing can incur damage to 20% of molecules, and the molecule-by-molecule distribution of these lesions is more heterogeneous than a Poisson distribution. Photodamage from UV shadowing is thus likely a widespread but unappreciated cause of artifactual heterogeneity in quantitative and single-molecule-resolution RNA biophysical measurements.

**INTRODUCTION**

As studies of RNA behavior seek greater quantitative precision and explanatory power, experimenters are re-investigating unusual features that were largely ignored in previous work. Multiple folding pathways, molecular individuality, highly heterogeneous kinetics, and long-lived metastable states are being uncovered for numerous RNA and RNA/protein systems – especially through single-molecule approaches – and have important implications for fully understanding the biological and evolutionary behavior of RNA [see, e.g., references (Zhuang et al. 2002; Tan et al. 2003; Xie et al. 2004; Lemay et al. 2006; Korennykh et al. 2007; Huang et al. 2009; Karunatilaka et al. 2010; Solomatin et al. 2010; Neupane et al. 2011)]. Nevertheless, few mechanisms for these molecule-to-molecule variations have been established, and there remains a concern, in some cases confirmed (Greenfeld et al. 2011), that some of the observed variations stem from impurities or misfolding arising during synthesis or subsequent handling (Pereira et al. 2010; Solomatin et al. 2011). Achieving a precise mechanistic understanding of RNA behavior will require a high degree of confidence that experimental RNA samples are pure and chemically well-defined.

One point of entry for chemical damage could be polyacrylamide-gel-based purification of RNA preparations, which is routine in biochemistry labs. In particular, purification protocols typically involve localizing samples in gels by illumination with hand-held ultraviolet (UV) lamps, which results in shadows on fluorescent screens. UV-induced lesions in nucleic acids, particularly photo-dimerization of proximal pyrimidines to produce cyclobutane-like or other linkages (Figure 1A), have been studied for decades due to their adverse biological and medical effects on genomic DNA (Greenstock et al. 1967; Sinha and Hader 2002). To avoid such damage during routine nucleic acid purification, researchers could load side-by-side replicate samples onto gels, some of which are 'sacrificed' to UV shadowing so that the other un-shadowed samples can be approximately located and then excised without risk of UV exposure.

Nevertheless, published protocols and handbooks for UV shadowing do not prescribe such precautions and lack recommendations for reasonable exposure times (Hendry and Hannan 1996; Clarke 1999; Andrus and Kuimelis 2001; Hartmann et al. 2005; Golden 2007). Consequently, many researchers may shadow their desired samples directly and risk UV-induced lesions. Furthermore, such lesions connecting sequence-adjacent residues do not change an RNA's mass and negligibly change its gel mobility. Thus, UV damage is not readily detected in downstream experiments involving mass spectroscopy or gel electrophoresis. Even nuclease-based tests (Sinha and Hader 2002; Greenfeld et al. 2011). have not given the sequence dependence or time-course of the damage in RNAs of interest. Due to this lack of quantitative assessment, it is possible that many researchers are unknowingly introducing a high frequency of lesions into their RNA preparations during gel purification.

Here, we report single-nucleotide-resolution experiments that quantify RNA damage incurred during UV shadowing for numerous RNAs, several commercially available UV lamps, and a series of exposure times. Our results strongly suggest that covalent damage – affecting 20% or greater of purified molecules and with unanticipated molecule-by-molecule distributions – are occurring in studies with gel-purified RNA molecules. It is thus important not only to avoid UV exposure in current RNA preparations but also to revisit systems in which non-intuitive results, such as heterogeneity of folding behavior, have been observed.

**RESULTS**
After establishing the presence of UV-induced damage in a wide range of gel-purified RNAs, we examine various factors that might mitigate such damage, from sequence to choice of lamps and filters to time of exposure.

*Survey of several RNAs confirms damage at pyrimidine-pyrimidine sites*
In initial studies by our laboratory investigating chemical modification of RNAs (Kladwang et al. 2011a; Kladwang et al. 2011c), we found that lesions from exposure to ultraviolet lamps were readily detected by reverse transcription with a fluorescent primer, a rapid readout for chemical structure mapping or footprinting experiments. Capillary electrophoresis of the reverse-transcribed DNA fragments provides a quantitative single-nucleotide-resolution assay for damaged positions.

Figure 1B presents the results of this procedure on seven RNA molecules visualized by UV shadowing after denaturing polyacrylamide gel electrophoresis (PAGE; 8% polyacrylamide with 7 M urea and 1x TBE electrophoresis buffer, 89 mM Tris-borate, pH 8.3; 1 mM EDTA). The seven sequences include riboswitch ligand-binding domains (adenine riboswitch from *Vibrio vulnificus*; a cyclic diGMP riboswitch from *Vibrio cholerae*; and a glycine-binding domain from *Fusobacterium nucleatum*); other natural RNAs (the P4-P6 domain of the *Tetrahymena* ribozyme, unmodified tRNA$^{phe}$ and 5S rRNA from *Escherichia coli*); and an artificial hairpin construct, the Medloop RNA (Kladwang et al. 2011a; Kladwang et al. 2011c). The samples were exposed to UV light with a hand-held lamp (Ultraviolet Products UVG-54, 254 nm, 6 W) at a distance of 10 to 20 cm from the sample. In all seven cases, samples exposed to the lamp for at least one second showed striking patterns of reverse transcriptase stops compared to samples that were excised from the same gels without UV exposure. For illustration, the data in Figure 1B are for samples exposed for 100 seconds, although damage was readily detected with smaller exposure times; see below for time-courses and quantification.

To investigate whether PAGE produces other covalent effects aside from UV-induced lesions, we also carried out control measurements with RNAs purified by non-gel means (phenol/chloroform extraction and gel filtration; MagMax RNA-binding beads; and hybridization of 3´ ends to complementary DNA oligonucleotides bound to magnetic beads). These experiments revealed no reproducible effects from UV-free PAGE purification aside from modest damage traced to oxidation from ammonium persulfate used to polymerize acrylamide (0.02-0.04% of RNA molecules affected); see SI Figure S1.

UV damage from photodimerization is expected to occur predominantly at sequence segments with neighboring pyrimidines (YY = UU, CU, UC, and UU) (Greenstock et al. 1967; Sinha and Hader 2002). To test this expectation, we assigned the sequences of the capillary electrophoretic traces by repeating the reverse transcription with 2´-3´-dideoxyguanosine triphosphate (ddGTP) to obtain reference peaks whose positions corresponded to C positions (see also below). Sequence annotation of the electropherograms shows that UV-induced damage indeed occurs predominantly at YY positions (boldface in Figure 1B).

To further confirm the sensitivity of general RNA sequences to UV damage, we designed five artificial RNA sequences U1 to U5 containing pyrimidine-pyrimidine sites arrayed in a pattern spelling out a warning message, "UV BAD" (Figure 1C). As predicted, after these RNAs were purified with UV shadowing and subjected to reverse transcription, the resulting capillary electropherograms displayed the expected message (Figure 1D).

*Detailed Sequence Dependence*
Given the visually striking prevalence of pyrimidine-pyrimidine sequences at UV-induced stops, we investigated whether the extent of damage could be quantitatively correlated with sequence. We assessed the degree of uniformity of the reactivity across pyrimidine-pyrimidine (YY) positions and whether damage occurs at other sites by quantifying capillary electrophoretic traces at single-nucleotide resolution using the HiTRACE software (Yoon et al. 2011).

As shown in Figures 1E & 1F, UV reactivities across all studied RNAs are not strongly stereotyped. For example, the distribution at UU positions is not strongly peaked; it spans a ten-fold range of reactivities and is approximately flat within this range. One explanation for this heterogeneity is residual structure in the RNA within the polyacrylamide gel despite the presence of 7 M urea; UV experiments in non-denaturing conditions often show modulation or enhancement of photodimerization due to secondary or tertiary structure formation (data not shown; see also, e.g., reference (Sawa and Abelson 1992)). Regardless of its origin, the spread in reactivities even for specific sequences currently precludes a strongly predictive quantitative model for UV damage from sequence.

The quantification of reactivities also showed that many segments of the RNA sequences that contained a pyrimidine after a purine, such as GU or AU, were covalently modified by UV exposure. Less reactivity was observed at other positions. See Figure 1E for sample histograms of AU vs. UU and AA; and Figure 1F for summary of median and ranges over all sequence doublets. Again, a possible explanation for effects at non-YY sequences is residual structure, e.g., base pairs or transient contacts that enable photodimerization of pyrimidines distant in sequence from one another. Alternatively or in addition, these positions may correspond to UV-induced lesions that are not the common cyclobutane-type photodimers (Figure 1A) (Ravanat et al. 2001; Sinha and Hader 2002). The isolation and characterization of these lesions through nuclease digestion and mass spectrometry may reveal further information on the damage but is currently challenging; in the following instead we continue to focus on practical considerations of how photodamage might be avoided.

*Different lamps*
It is well known that photodimerization of pyrimidines is promoted by UV radiation at shorter wavelengths (Saitou and Hieda 1994) and so might be avoided by using UV lamps mainly emitting radiation at longer wavelengths and/or filtering out shorter wavelengths. We surveyed the Stanford Biochemistry Department for lamps with different emission wavelengths that were in use for UV fluorescence/absorption visualization and tested their effects in localizing the P4-P6 RNA during gel purification. All lamps were manufactured by Ultraviolet Products, Inc. (UVP).

Lamps emitting longer wavelength radiation (UVP 3UV-34 three-setting lamp with 302 nm and 365 nm settings; and UVP UVL-56, 366 nm) gave weak or no detectable damage on the RNA (Figure 1G), but they also did not give detectable UV shadows with the amounts of RNA used (50–100 μg). The other lamps (two UVP UVG-54 lamps, 254 nm) all gave easily detectable shadows but also clear UV-induced damage bands in the reverse transcription assay (Figure 1G). Note that the lamps recommended by several published protocols (Hendry and Hannan 1996; Andrus and Kuimelis 2001; Hartmann et al. 2005) as well as the UVP website for shadowing and nucleic acid purification are the short-wavelength (254 nm) emitters, which we found to cause the most damage.

*Time-course of UV damage*

Perhaps the most obvious route to reducing covalent RNA damage during UV shadowing would be to reduce the UV dose illuminating the sample. This reduction can be accomplished in two ways. Moving the source further from the sample diminishes UV exposure, but risks reducing the clarity of the RNA band's UV shadow and precluding visual detection. A more effective strategy to diminish UV exposure might be to reduce the exposure time to the minimum required for finding and marking the desired band. We therefore sought to measure the rate of UV damage by carrying out a time-course of UV exposure. We note that the minimal scanning and marking time is approximately 5 seconds, with some protocols suggesting exposures up to 30 seconds (Hendry and Hannan 1996). We therefore exposed gel samples of the P4-P6 RNA to UV lamp radiation from seconds to several minutes (to obtain an endpoint) and quantified UV-induced damage as above.

The results of this timecourse are shown in Figure 1H. For the earliest time points, the UV damage pattern is detectable above background within 3 seconds, suggesting that significant damage is unavoidable in this procedure even for fast exposure times. Such damage within seconds was seen for experiments with the lamp placed 5 cm, 10 cm, and 20 cm from the sample. For time points between 20 seconds up to 600 seconds, the data approaches an apparent 'steady-state' pattern. At the longest time points (40 and 60 minutes) the reverse transcription signal is sharply attenuated for extension lengths beyond 20 to 30 nucleotides, which corresponds to a long tract of pyrimidines in the P4-P6 sequence getting saturated with lesions.

Quantifying these data required taking into account an unusual feature of the covalent modification process. Since the very goal of UV shadowing is to visualize the RNA, the RNAs must absorb a substantial fraction of the radiation as it passes through the gel. Thus, RNAs at the 'top' of the gel slice (closest to the UV lamp) will receive substantially higher exposure than those at the bottom, where the radiation has been attenuated by absorption. Indeed in the limit of high RNA concentrations, only the very top layer is damaged, which we term a 'skin effect', by analogy to the exponential reduction of radiation that penetrates into conductors in electromagnetism theory (Jackson 1975).

Because different populations of RNAs receive different exposures, a Poisson process with a single rate of modification does not hold. Instead, the UV damage rate is attenuated exponentially with 'skin depth' $\lambda$ as a function of the depth through the gel slice. The resulting model leads to excellent fits of the experimental measurements [Figure 2B; see equations (1)-(4)], with $\lambda = 0.12 \pm 0.05$ mm and with the damage rate at the top of the gel as $k = 0.03 \pm 0.01$ s$^{-1}$ (summed over all nucleotides of the RNA). Independent measurements that used small concentrations of 2´-3´-dideoxyguanosine triphosphate (ddGTP) to give 'internal control' bands in the reverse transcription pattern (SI Figure 2 and SI Text) also fit well to the skin effect model with best-fit parameters $\lambda = 0.10 \pm 0.05$ and $k = 0.039 \pm 0.005$ s$^{-1}$, in excellent agreement with the measurement without ddGTP. A simple Poisson model (no skin effect) gave significantly worse fits to these data and to additional data varying RNA concentration (see SI Figure 2 and SI Text).

The estimated skin depth $\lambda$ (0.12 ± 0.05 mm) is less than the gel thickness 0.5 mm, as expected for a population absorbing enough radiation to generate a visible UV shadow on an underlying fluorescent screen. Furthermore, assuming the extinction coefficient for absorption by denatured RNA of 25–50 mL mg$^{-1}$ cm$^{-1}$, this $\lambda$ corresponds to RNA at a concentration of 1–3 mg/mL, which accords well with a value of 2–4 mg/mL computed from the loaded amount of RNA (50–100 μg) and the approximate dimensions of the band in the gel (1.0 cm x 0.5 cm x 0.5 mm). We note that, in principle, photodimerization rates vary with UV wavelength; radiation at different wavelengths may penetrate with different skin depths; and polyacrylamide may also possess a complex UV absorption spectrum. However, the data do not require these more complex extensions to the skin effect model to achieve reasonable fits. We have also expanded the model to include diffusion of RNA molecules within the gel but this extension did not improve the fits (not shown).

The skin effect model further provides an intuitive explanation for the apparent 'steady state' behavior of the UV modification pattern observed from 10 seconds to 600 seconds (Figures 2D and 2E). At early timepoints, the only damaged RNA molecules are those closest to the lamp, at the top of the gel slice (Figure 2F). At later timepoints, this very top layer is so damaged that its RNAs do not contribute to the observed UV photodimer pattern read out by reverse transcription. RNA molecules at the very bottom of the gel also do not contribute to the damage pattern, but for a different reason – they are unmodified due to protection from the top layer. Instead the observed pattern of reverse transcription products is due to an intermediate 'active' layer with skin depth $\sim\lambda$ (Figure 2G). This active layer travels down the gel slice slowly; its starting depth increases logarithmically with time (Figure 2H) while its thickness stays nearly constant at $\lambda$. The RNAs in this active layer contribute a constant UV damage pattern, and this steady-state pattern is readily derived [Methods eq. (5)]. Once the active layer reaches the bottom of the gel slice, all the RNAs in the gel have succumbed to numerous damage events, and their reverse transcription is strongly attenuated (Figure 2I).

**DISCUSSION**

*UV damage may be ubiquitous in UV-shadowed RNA preparations*

UV damage of nucleic acids, despite being well appreciated as a ubiquitous biological phenomenon and a known risk during gel purification of RNA, is not always controlled or quantitatively assessed during *in vitro* RNA investigations. This issue may be exacerbated by a previously untested assumption that short exposure times give negligible damage and to the difficulty of directly detecting the photodimer products within long RNAs. We have now quantified this UV-induced damage using reverse transcription read out by capillary electrophoresis. Across several natural and artificial RNAs, across multiple sequence contexts (dominantly at but not limited to pyrimidine doublets), and with multiple lamps that are recommended for UV shadowing, we have detected substantial RNA damage occurring in routine UV shadowing protocols.

*Quantifying the extent and molecule-by-molecule heterogeneity of damage*

To more precisely assess the UV-induced damage, we have quantified and modeled the lesions occuring within polyacrylamide gels versus time. We found surprising results for the damage rate and the molecule-by-molecule distribution of lesions.

First, the time course of UV damage presents a long phase in which the reverse transcription pattern reaches an apparent steady state, and a simple Poisson model cannot fit this behavior quantitatively. Instead, these data, as well as independent experiments with RNAs in solution, are consistent with a 'skin effect' model: RNA molecules closest to the lamp are damaged at a higher rate than molecules further below, which are protected by attenuation of the UV radiation by absorption. Fits to this model give a rate of modification for the P4-P6 RNA (202 nucleotides) of 0.03–0.04 $s^{-1}$ at the top of the gel, attenuated exponentially with a skin depth of $\lambda = 0.10$-$0.12$ mm under the tested conditions. Properties of the RNA damage distribution can be integrated from the skin effect model, as follows.

If a researcher spends 10 to 20 seconds to locate and mark an RNA gel band by UV shadowing, the fraction of gel-excised molecules incurring at least one lesion is 0.10 to 0.17; and the mean number of lesions per RNA is 0.16 to 0.27. For comparison, numerous single-molecule folding and catalysis studies have uncovered sub-populations of kinetically distinct molecules at frequencies of 10-30%. We note that a number of factors could increase the fraction of damaged molecules to 50% or higher. For lower concentrations of RNA, the overall lesion rate will increase since the skin depth will be longer. Further, lower RNA concentrations will result in weaker UV shadows that require longer exposure times – and concomitantly higher damage rates – to permit confident localization and marking of the band. Finally, the lesion rate per RNA increases with the length of nucleotides of the molecule, so that preparations of, e.g., the *Tetrahymena* group I ribozyme (388 nts) under our tested conditions would contain 50% of molecules with at least one lesion after 20 seconds of UV shadowing. Strategies to reduce these damage rates are further discussed in the next section.

A second major insight from these quantitative measurements is that a population of UV-shadowed RNAs is, on a molecule-by-molecule basis, quite different in its distribution of damage than would be predicted from a simple Poisson model. The heterogeneity of the number of lesions is significantly larger. Figure 2J illustrates the fraction of molecules containing 0, 1, 2, ... lesions for RNA populations with average lesion number of 1.0 (seen at 60 seconds of UV exposure under the tested conditions in Figure 2). The UV-shadowed population gives a larger spread in the number of lesions than the Poisson model. For example, the variance in the lesion rate in the skin effect model is 3.8, substantially greater than the variance in the Poisson model (1.0). In addition to heterogeneity in lesion number, there is the heterogeneity in lesion location – the damage can occur nearly anywhere in the molecule (Fig. 1B-1F). These heterogeneities appear to have been unappreciated in prior work using UV-visualized RNA preparations.

*Assessing and avoiding damage*
All manipulations of a nucleic acid after its synthesis can lead to covalent damage. In many cases, such as cleavage from nuclease contaminants, analytical electrophoresis or mass spectrometry of samples in downstream experimental steps can give readouts of such damage. However, these methods are not directly sensitive to UV-induced photodimers. Based on the results herein, we advocate using reverse transcription and capillary electrophoresis – which can be carried out in an afternoon on tens of samples – to rapidly assess UV photodimer formation and other undesired covalent modifications.

More generally, it appears prudent to avoid UV exposure to critical samples. Reducing exposure time to under one second, increasing lamp distance beyond 10 cm, or loading higher concentrations of RNA into thicker gels should diminish UV damage to most of the population but will still leave heterogeneities due to the skin effect. It is perhaps safest to sacrifice loaded samples (or an edge of the gel band of interest) to enable localization of nearby RNA that remains undamaged. Nevertheless, other sources of damage can occur during gel purification, even without UV treatment. For example, heating prior to or during electrophoresis may cause some form of covalent damage (Greenfeld et al. 2011). Furthermore, we have detected oxidative modifications at a rate of 0.02-0.04% per RNA that can be traced to ammonium persulfate used to polymerize acrylamide gels (SI Fig. 1) (Chiari et al. 1992). This damage measurement may be an underestimate if the reverse transcriptase used herein (SuperScript III) bypasses oxidative lesions at a non-negligible rate.

We conclude that the best strategies for RNA purification for biophysical measurements should avoid gel electrophoresis altogether. Several groups have proposed non-gel-based purification methods for crystallography, NMR, and high-throughput biophysics applications [see, e.g., references (Kieft and Batey 2004; Lukavsky and Puglisi 2004; Batey and Kieft 2007; Kladwang and Das 2010; Kladwang et al. 2011b)], and these methods either give sub-second UV doses (during chromatographic detection) or avoid UV exposure. Taken together with unsafe UV visualization methods in published protocols, our data indicate that UV shadowing is likely causing unappreciated covalent damage in numerous high-precision and single-molecule-resolution RNA biophysical measurements. It is important to revisit these experiments with RNA samples prepared with gel-free and UV-free purification methods.

**ACKNOWLEDGEMENT**
We thank M. Greenfeld & D. Herschlag for useful discussions and sharing of their manuscript before publication; the Stanford Biochemistry Department for sharing ultraviolet lamps; and M. Zhang and H. Mabuchi for use of a hand-held UV detector. This work was supported by the Burroughs-Wellcome Foundation (CASI to RD).

**SUPPORTING INFORMATION PROVIDED**
Figure showing oxidative damage from ammonium-persulfate-polymerized gels; and text and figure describing additional tests of the skin effect model versus the simple Poisson model.

**EXPERIMENTAL SECTION**
*Polyacrylamide gel electrophoresis and UV shadowing*

All RNAs were prepared as in prior work (Kladwang et al. 2011a; Kladwang et al. 2011b) by *in vitro* transcription with T7 RNA polymerase from PCR products (with the 20 bp T7 promoter sequence TTCTAATACGACTCACTATA included at the 5´ end), with transcription volumes up to 1.5 mL. Transcriptions were precipitated by adding 1/10 volume of sodium acetate (pH 5) and 3 volumes of cold ethanol (taken out of 4 °C storage), cooling on dry ice for at least 15 minutes, and centrifuging at 14,000 g for 1.5 hours. After removal of supernatant, pellets were rinsed with 1 mL 70 % cold ethanol twice, dried in air for at least 30 minutes, and resolubilized in deionized water at volumes equal to 1/10 of the original transcription. A half volume of denaturing loading buffer (90% formamide, 0.1% xylene cyanol, 0.1% bromophenol blue) was added, and the samples were loaded onto polyacrylamide gels. The gels were 0.5 mm in thickness, 20 cm in height (direction of electrophoresis), and 27 cm in width. The gel mix contained 1x TBE (89 mM Tris-Borate, 1 mM EDTA), 8% polyacrylamide (29:1 acrylamide:bis, Sigma), and 7 M urea, and were polymerized by the addition of 1/100 volume of 10% ammonium persulfate and 1/1000 volume of TEMED (*N*,*N*,*N'*,*N'*-Tetramethylethylenediamine); after pouring between glass plates, the gels were allowed at least 1.5 hours to polymerize. Variations with longer polymerization times and use of flavin mononucleotide as the polymerizing reagent are discussed in SI Fig. S1. Gels were run at 25 W or less for 1 to 3 hours (temperatures remained less than 40 °C under electrophoresis conditions).

Gels were transferred from gel plates onto UV-transparent plastic wrap (Saran), covered with wrap on both sides, and placed on a fluorescent intensifying screen (Dupont Cronex). Samples were exposed to UV hand-held lamps (Ultraviolet Products UVG-54, 254 nm, 6 W; unless specified otherwise) and boxes were marked on plastic wrap around band locations with Sharpie markers; samples not undergoing shadowing were covered with aluminum foil. For time course measurements (Fig. 2), early timepoints were acquired by turning on the lamp for a few seconds (for warm-up) and transiently removing the foil for the presented times. Gel slices were excised with sterile, disposable scalpels (BD) after peeling back plastic wrap and placed in 1.5 mL Eppendorf tubes with 200 µL deionized water. RNAs passively eluted into the water during incubation overnight at 4 °C, and concentrations were estimated by absorption measurements at 260 nm on a Nanodrop spectrophotometer.

*Reverse transcription*

RNA sequences are presented in Fig. 1; all sequences included an additional 20 nt sequence AAAGAAACAACAACAACAAC at the 3´ end as a common reverse transcription binding site. Reverse transcription with Superscript III (Invitrogen) and capillary electrophoresis on ABI 3100 and 3730 machines were carried out as previously described, using poly(A) purist magnetic beads (Life Technologies) to accelerate purification steps (Kladwang et al. 2011a; Kladwang et al. 2011c; Kladwang et al. 2011b). For reverse transcriptions with 'doping' from 2´-3´-dideoxyguanosine triphosphate (ddGTP), concentrations of nucleotides were 0.05 mM ddGTP and 1 mM dATP, dCTP, dTTP, and dITP (2´-deoxyinosine triphosphate). Data were aligned and quantitated with the HITRACE software (Yoon et al. 2011).

*Model for UV damage timecourses*

Data for ultraviolet timecourses were fit to an extension of the basic Poisson model, in which modifications at different sites occur independently and stochastically [see, e.g., references (Aviran et al. 2011; Kladwang et al. 2011c) for prior applications to chemical mapping data]. Assume that the total rate of modification at all sites is $k$, and the fractional modification rate each site $i$ is given as $k_i = \alpha_i k$. The sum over the fractional modification rates $\alpha_i$ is unity. The skin effect model is different from the simple Poisson model in that the damage rates are not constant for all RNAs but are attenuated at depth $z$ with characteristic length $\lambda$:

$$m_i(z) = \alpha_i k e^{-z/\lambda} \qquad (1)$$

The lesion probability at site $i$ at time $t$ is then:

$$p_i(z,t) = 1 - e^{-m_i(z)t - b_i} \qquad (2)$$

where $b_i$ is the (UV-independent) background rate of stopping reverse transcription. The fraction of molecules $F_n$ with $n$ lesions is given by the Poisson formula, but averaged through different depths:

$$F_n(t) = \frac{1}{L}\int_0^L \frac{1}{n!}e^{-p(z,t)}p^n(z,t)dz \qquad (3)$$

The fraction of reverse transcription product $f_i$ at site $i$ after traversal through sites 1, 2, … $i-1$ is:

$$f_i(t) = \frac{1}{L}\int_0^L p_i(z,t)\prod_{j=1}^{i-1}\left[1-p_j(z,t)\right]dz, \qquad (4)$$

Note that $i$ indexes the position in the reverse transcribed cDNA (5´ to 3´), with $i = 1, 2,$ … corresponding to primers extended by one, two, … nucleotides. The equations (1)-(3) were computed in MATLAB (Mathworks), with the integrals numerically approximated by subdividing the gel slice into subslices of length 0.001 $L$. The site-dependent background stopping probability $b_i$ was measured by the reverse transcription experiment without UV exposure. The site-specific UV-damage rate $\alpha_i$ was estimated based on measurements with 600 seconds of UV exposure, subtracting $b_i$ and normalizing to unity. Similar parameter fits were obtained for $\alpha_i$ estimated from later timepoints and also by iterating the fits so that new $\alpha_i$ were obtained based on the ratio of data at late timepoints with predictions based on the original estimates of $\alpha_i$. Data at the very 5´ and 3´ ends gave strong signals that saturated the capillary electrophoresis fluorescence detector and were not used in the fits (see Fig. 2); so the total modification rates given in the text are slight underestimates. Note that a standard Poisson model can be obtained from equations (1)-(4) above by setting $\lambda \to \infty$ (no attenuation of radiation by absorption).

To model measurements that included ddGTP for internal controls, the 'background' stopping probabilities $b_i$ were adjusted based on reverse transcription measurements without UV but including ddGTP. Suppose that there is a ddGTP signal nucleotide at $h$, and the next UV-induced signal is at $i>h$, such that there is no cross contamination ($k_h = 0$, $b_i = 0$, and no bands in between $h$ and $i$); examples of these pairs are shown in SI Figures 2D and 2E. For the standard Poisson model, the ratio of the two signals is given by $p_i(1-p_h)/p_h \approx p_i/b_h$ with the complex attenuation factor $\prod_{j=1}^{h-1}(1-p_h)$ canceling out.

For times up to saturation, this ratio increases linearly with time since $b_h$ is constant and $p_i$ increases as in eq. (2), which is linear for $p_i < 1$. For the general skin effect model, the ratio of the signals is more complex but is numerically calculated as above.

For a large range of times and negligible background $b_i$, the reverse transcription pattern [equation (4)] reaches an approximate 'steady state' in which the top of the gel slice is saturated with lesions, the bottom of the gel slice to remain undamaged, and an intermediate subslice of RNAs ('active' layer) dominates the contribution to the reverse transcription signal. The steady state is given by the simple formula:

$$f_i = \frac{\sqrt{2\pi}\lambda}{eL} \frac{\alpha_i}{\sum_{j=1}^{i-1}\alpha_i} \quad \text{(independent of time } t\text{)} \tag{8}$$

The equation is derived in the Supporting Text.

**FIGURE LEGENDS**

**Figure 1. Characterizing RNA damage during UV shadowing by reverse transcription read out by capillary electrophoresis.** **(A)** Cyclobutane dimerization product of two uracil bases initiated by UV absorption. **(B)** Capillary electropherograms of reverse-transcription products for seven RNAs that were or were not UV-shadowed during polyacrylamide gel purification. Sequence is as given except for reverse transcription binding site at 3´ end (see Methods). Boldface letters and asterisks mark pyrimidine-pyrimidine doublets. Exposure time was 100 seconds by a hand-held UVG-54 lamp (Ultraviolet Products) at 10 cm distance to samples. **(C)** Design of sequences U1 to U5 to confirm damage predominantly at pyrimidine-pyrimidine (UU, UC, CU, and CC) sites. **(D)** Capillary electropherograms of gel-purified sequences U1 to U5. **(E)** Histograms of UV reactivities at AA, AU, and UU sites from RNAs in (B); reactivities are relative to mean UV reactivity seen for each construct. **(F)** Box plot of reactivities across all dinucleotide types, shown as medians (white symbols), $25^{th}$–$75^{th}$ percentile (interquartile) ranges (black boxes), most extreme data points that are outside the interquartile range by no more than 1.5 times this range (whiskers), and values beyond the whisker range (small black symbols). **(G)** Effects of different UV handheld lamps on the P4-P6 RNA (P4-P6 domain of the *Tetrahymena* group I ribozyme). **(H)** Time-course ultraviolet damage for the P4-P6 RNA shadowed during gel purification by a UVG-54 lamp at a distance of 5 cm.

**Figure 2. Time-course of UV damage and a 'skin effect model'.** **(A)** Quantitation of reverse transcribed products for the P4-P6 RNA exposed to increasing ultraviolet doses (data correspond to Figure 1H). **(B)** Fit of data in (A) to a skin effect model in which the UV lesion rate in the 0.5 mm gel slice is 0.032 $s^{-1}$ but attenuated exponentially by absorption with skin depth λ = 0.1 mm. **(C), (D), (E),** and **(F):** Predictions of the skin effect model for the distribution of damage (here, the fraction of RNAs with at least one lesion) at different depths of the gel slice (vertical axis) at different points in the observed time-course. **(G)** Molecule-by-molecule distribution of lesions in the Poisson and skin effect models when the average number of lesions per molecule is 1 [corresponding to the 60 second timepoint (D)].

**Figure 1**

**Figure 2**

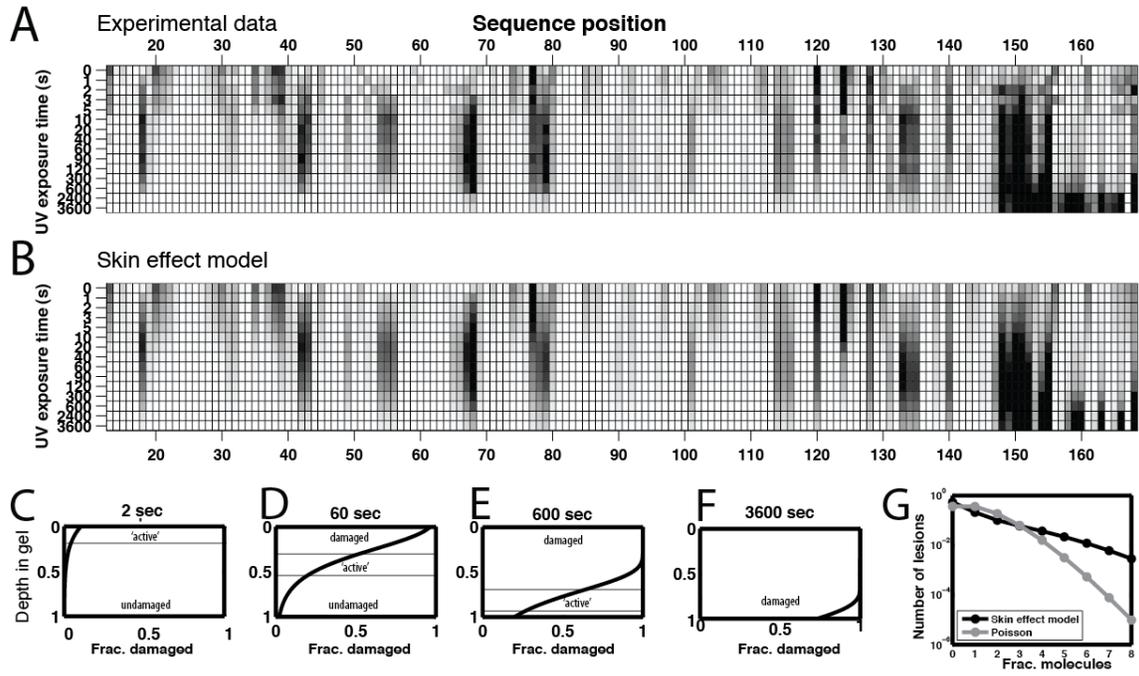

# Supporting Information for "Ultraviolet Shadowing of RNA causes Substantial Non-Poissonian Chemical Damage in Seconds"

Wipapat Kladwang[1], Justine Hum[1], and Rhiju Das[1,2*]

Department of Biochemistry[1] and Department of Physics[3], Stanford University, Stanford, California 94305

[*] To whom correspondence should be addressed. Phone: (650) 723-5976. Fax: (650) 723-6783. E-mail: rhiju@stanford.edu.

This supporting information contains:

*Supporting Text S1. Inaccuracy of the simple Poisson model*

*Supporting Text S2. Skin effect model fits to ddGTP doping experiments*

*Supporting Text S3. RNA concentration test of skin effect model*

*Supporting Text S4. Derivation of approximate steady state reverse transcription pattern for UV exposure timecourses*

*Supporting Figure S1. Evidence for oxidate damage from PAGE purification.*

*Supporting Figure S2. Discrimination of Poisson model from 'skin effect' model.*

*References for Supporting Information*

**Supporting Text**

*S1. Inaccuracy of the simple Poisson model*

We first analyzed the UV damage timecourse (main text Figures 1H and 2A) with a simple Poisson model that assumes that the UV dose to all molecules is uniform rather than attenuated with depth. Thus, the lesion probability at any given site should increase linearly with time with some site-dependent rate constant $k_i$. In this model, the total number of lesions per molecule increases with rate $k = \sum_i k_i$, and the molecule-by-molecule distribution of lesions follows the Poisson formula (see Eqs. 1-4 in Methods). The observed reverse transcription stops should rise at all positions, up to a timescale $\tau \sim 1/k$. At that point, the number of average lesions per molecule exceeds unity, and at longer times $t > \tau$, we should observe significant attenuation of long reverse transcription products. Upon optimizing over the site-dependent rates, the model gives approximate agreement with the observed data, but the fit is quantitatively unsatisfactory (cf. SI Figures 2A and 2B). Instead of saturating and then dropping around some timescale $\tau$, the pattern of damage remains invariant over approximately two orders of magnitude in time, between 10 seconds to 600 seconds (10 minutes) and this effect cannot be fit by the Poisson model.

A more stringent falsification of the Poisson model involves reverse transcriptions that are doped with 2´-3´-dideoxyguanosine triphosphate (ddGTP). This doping gives internal controls at C positions, and the resulting bands' intensities reflect a constant reverse transcription stopping rate, which should allow the attenuation of reverse transcription for UV damage fragments to be estimated. In the simple Poisson model, positions with UV-dependent damage should increase linearly with time compared to the intensities of these ddGTP-induced reference signals (see Methods). The ratio of any UV signal to a nearby ddGTP control band should thus increase dramatically, shown as gray curves in SI Figures 2D and 2E. Instead, the UV damage signals (black symbols) approach an apparent steady-state from 10 seconds to 600 seconds, during which their intensity stays approximately constant relative to the ddGTP internal control signals. Other strategies to estimate damage rates, e.g., based on the fraction of unmodified products remaining after UV damage, also give poor fits (not shown).

These observations indicate that UV-induced damage can not be understood in terms of the simplest Poissonian model for chemical modification, and accurate quantification of the damage extent requires the skin effect model, discussed in the main text and below.

*S2. Skin effect model fits to ddGTP doping experiments*

As an additional test of the skin effect model, we fitted the reverse transcription measurements using ddGTP doping to provide reference signals (SI Figure 2C). Interestingly, in the context of the skin effect model, the ddGTP signals are mostly contributed by undamaged RNAs at the bottom of the gel (main text Figure 2E), a different population than those contributing the UV signals in the 'active' layer, so they do not truly provide 'internal' controls. Nevertheless, the mathematical description of the process is straightforward [eqs. (1)–(4) in Methods], and the observed data are fit well to the this model (see black lines in SI Figures 2D and 2E) with skin depth $0.10 \pm 0.05$ mm and total modification rate of $0.039 \pm 0.005$ s$^{-1}$, in agreement with fits to data without ddGTP doping presented in the main text.

*S3. RNA concentration test of skin effect model*

A straightforward qualitative prediction of the skin effect model is that higher concentrations of RNA should protect the overall population from UV-induced damage, since the the radiation will penetrate less into the population (shorter skin depth). Because it is difficult to control RNA concentrations during gel electrophoresis, we tested this prediction with solution measurements. We prepared P4-P6 RNA with concentration of 0.05 mg/mL and 0.5 mg/mL within wells of depth 1.5 mm in 1X TBE electrophoresis running buffer (89 mM Tris-Borate, 1 mM EDTA). Replicates in 0.1X TBE gave indistinguishable results. After 60 seconds of UV treatment, the samples with lower RNA concentrations showed damage products (SI Figure S2F), while samples with the 10-fold higher RNA concentrations gave reverse transcription patterns identical to samples without UV exposure, confirming the predictions of the skin effect model.

*S4. Derivation of approximate steady state reverse transcription pattern for ultraviolet exposure timecourses*

Starting with equation (4) of the main text for the fraction of reverse transcription product $f_i$, we substitute eqs (1) & (2) [skin effect damage model] to yield

$$f_i = \frac{1}{L}\int_0^L \exp\left(-tke^{-z/\lambda}\sum_{j=1}^{i-1}\alpha_j\right)\left[1-\exp\left(-tke^{-z/\lambda}\alpha_i\right)\right]dz \qquad (S1)$$

We assume that $b_i$ is negligible. Now change variables, defining $q = kte^{-z/\lambda}$ and $dq = kte^{-z/\lambda}dz/\lambda = -qdz/\lambda$:

$$f_i = \frac{1}{L}\int_{kte^{-L/\lambda}}^{kt}\left[e^{-q\sum_{j=1}^{i-1}\alpha_j}\left(1-e^{-q\alpha_i}\right)\right]\frac{\lambda dq}{qL} \qquad (S2)$$

The parameter $q$ defines the UV dose within each gel subslice. In the skin effect model, for large enough $t$, $q$ ranges from undamaged ($kte^{-L/\lambda} \ll 1$) at the bottom of the gel to highly damaged ($kt \gg 1$) at the top of the gel ($z = 0$). The expression in square brackets has a maximum that is peaked as a function of $q$ for sites $i$ at least a few nucleotides from the reverse transcription start site 3´ end ($i \gg 1$). We seek to approximate the eq. (S2) by a Taylor expansion around this peak:

$$f_i = \frac{1}{L}\int_0^\infty e^{r|_{q_{max}} + \frac{1}{2}\left(\frac{\partial^2 r}{\partial q^2}\right)_{q_{max}}(q-q_{max})^2}\frac{\lambda dq}{q_{max}L} \qquad (S3)$$

where $r$ is the logarithm of the expression in square brackets in (S2):

$$r = \log\left[e^{-q\sum_{j=1}^{i-1}\alpha_j}\left(1-e^{-q\alpha_j}\right)\right] = -q\sum_{j=1}^{i-1}\alpha_j + \log\left(1-e^{-q\alpha_j}\right) \quad (S4)$$

Setting $\dfrac{\partial r}{\partial q} = 0$ gives the maximum of the expression at:

$$q_{max} = \frac{1}{\alpha_i}\log\left(1+\frac{\alpha_i}{\sum_{j=1}^{i-1}\alpha_j}\right) \approx \frac{1}{\sum_{j=1}^{i-1}\alpha_j} \quad (S5)$$

At this point,

$$r\big|_{q=q_{max}} = -1 + \log\left(\frac{\alpha_i}{\sum_{j=1}^{i-1}\alpha_j}\right) \quad (S6)$$

The second derivative of $r$ at $q_{max}$ evaluates to:

$$\frac{\partial^2 r}{\partial q^2}\bigg|_{q_{max}} = -\sum_{j=1}^{i-1}\alpha_j\sum_{j=1}^{i}\alpha_j \approx -\left(\sum_{j=1}^{i}\alpha_j\right)^2 \quad (S7)$$

The integral expression (S2) then becomes:

$$f_i = \frac{1}{L}\int_0^\infty \exp\left(-1+\log\left(\frac{\alpha_i}{\sum_{j=1}^{i-1}\alpha_i}\right) - \frac{1}{2}\left(\sum_{j=1}^{i}\alpha_j\right)^2(q-q_{max})^2\right)\frac{\lambda dq}{q_{max}L} \approx \frac{\sqrt{2\pi}\lambda}{eL}\frac{\alpha_i}{\sum_{j=1}^{i-1}\alpha_i}$$

It is independent of time – a steady-state solution – as long as we are at some time where the top of the gel has received a high UV dose; the bottom of the gel has received a low UV dose; and λ is small compared to the gel thickness.

**Supporting Figure S1. Evidence for oxidate damage from PAGE purification.** RNAs purified from polyacrylamide gels polymerized with ammonium persulfate (APS) show covalent damage even in the absence of UV shadowing (two replicates shown). Data shown are from capillary electrophoresis of reverse transcription products; strong bands correspond to covalent damage that can stop Superscript III reverse transcriptase. See marked bands for oxidative damage positions [mostly apparent at guanosines, likely due to formation of 8-oxo-guanosine or further oxidative products (Luo et al. 2000)]. The damage is substantially reduced if purification is carried out with gels that have been left overnight (12 hours compared to 1.5 hours) to polymerize after APS addition. The damage is reduced further in gels polymerized by photoactivated radical formation with flavin mononucleotide (FMN, also called riboflavin 5´-phosphate) present at 0.1% concentration and exposed on a light box for 12 hours (Chiari et al. 1992). Remaining bands in FMN-polymerized gel experiment appear to be due to reverse transcription stops at secondary structure and are also seen for freshly transcribed RNA purified by hybridization to oligonucleotide beads.

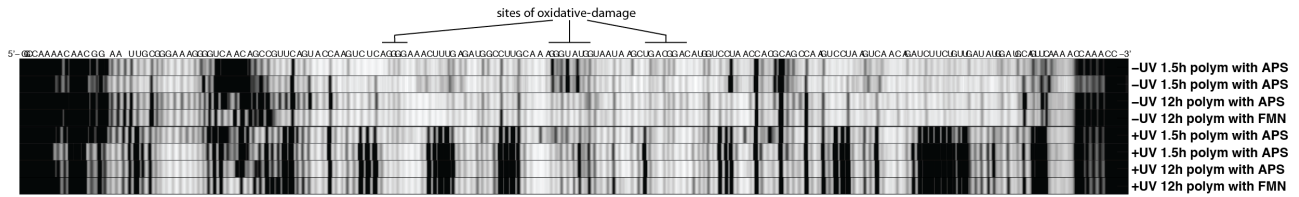

**Supporting Figure S2. Discrimination of Poisson model from 'skin effect' model** *(figure on next page)*. **(A)** Quantitation of reverse transcribed products for the P4-P6 RNA, exposed to ultraviolet damage from a UVG-54 lamp at a distance of 5 cm for different times; same data as main text Figure 2A. The bottom half shows additional data for the same RNA samples with 2´-3´-dideoxyguanosine triphosphate (ddGTP) included to produce 'internal control' bands at cytidine positions. **(B)** Fit of data in (A) to a Poisson model. The best-fit total modification rate per RNA ($k$) was 0.005 s$^{-1}$. **(C)** Fit of data in (A) to a skin effect model in which the UV lesion rate in the 0.5 mm gel slice is $k = 0.032$ s$^{-1}$ but attenuated exponentially by absorption with skin depth $\lambda = 0.1$ mm. **(D)** and **(E)**: Estimated damage at specific positions 78 (D) and 134 (E) based on the observed reverse transcription signal, corrected based on ddGTP internal control signals at 76 and 132, respectively. Gray and black curves represent predictions of the Poisson model (B) and skin effect model (C) for this observable. **(F)** Confirmation of prediction that higher molecular concentrations should protect the RNA population from damage; measurements were in solution with 1 mm liquid depths, UV exposure for 60 seconds, and RNA concentrations of 0.05 mg/mL and 0.5 mg/mL (marked 1x and 10x RNA). Buffer concentrations tested were 89 mM Tris-Borate, 1 mM EDTA (1x TBE) and 8.9 mM Tris-Borate, 0.1 mM EDTA (0.1X EDTA). Note that the samples did not include urea denaturants and the resulting partially folded RNAs gave less UV-induced damage than denaturing-PAGE-purified samples assayed elsewhere.

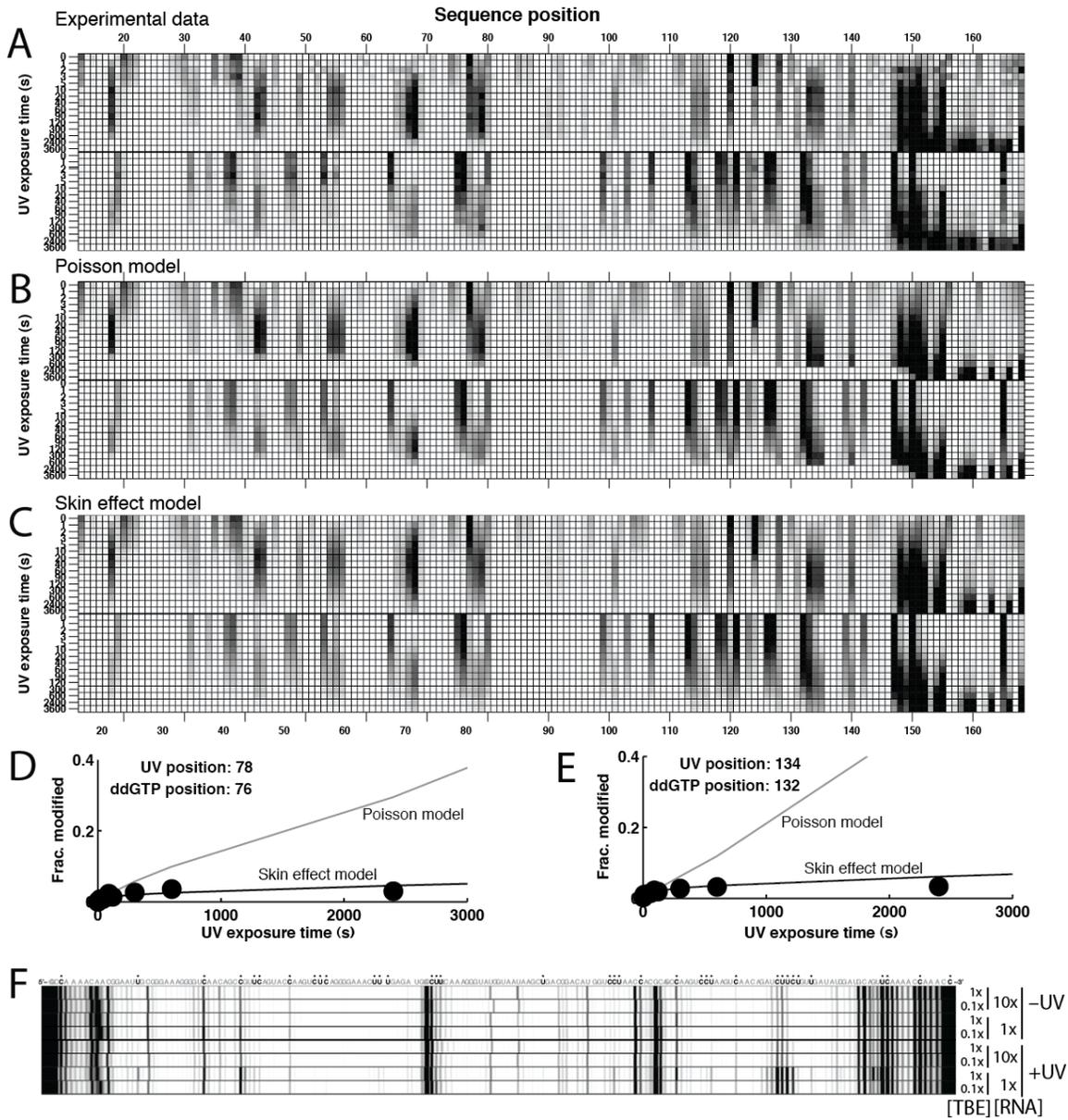

**References for Supporting Information**